\documentclass[aps,prc,twocolumn,nofootinbib,altaffilletter]{revtex4-1} 

\usepackage{epsfig}
\usepackage{color}
\usepackage[latin1]{inputenc}
\usepackage{float,amsmath}
\usepackage{graphicx}
\usepackage{comment}
\usepackage{ulem}
\usepackage{bbold}

\newcommand{\nc}{{N_\mathrm{c}}}
\newcommand{\id}{\mathbb{1}}
\newcommand{\tr}{\text{tr}}
\newcommand{\Nt}{N_{\bot}}
\newcommand{\at}{a_{\bot}}

\begin{document}

\title{The shape of the proton at high energies}
\author{S\"oren Schlichting}
\email{sschlichting@bnl.gov}
\affiliation{Physics Department, Brookhaven National Laboratory, Upton, NY 11973, USA}
\author{Bj\"orn Schenke}
\email{bschenke@bnl.gov}
\affiliation{Physics Department, Brookhaven National Laboratory, Upton, NY 11973, USA}

\begin{abstract}
We present first calculations of the fluctuating gluon distribution in a proton as a function of impact parameter and rapidity
employing the functional Langevin form of the JIMWLK renormalization group equation. 
We demonstrate that when including effects of confinement by screening the long range Coulomb field of the color charges, 
the evolution is unitary. The large-$x$ structure of the proton, characterized by the position of three valence quarks, 
retains an effect on the proton shape down to very small values of $x$. 
We determine the dipole scattering amplitude as a function of impact parameter and dipole size and extract the rapidity evolution of the
saturation scale and the proton radius.
\end{abstract}
  
\maketitle


\section{Introduction}
High energy hadronic and nuclear scattering experiments at the Relativistic Heavy-Ion Collider (RHIC) and the Large Hadron Collider (LHC), as well as deeply inelastic scattering (DIS) and exclusive diffractive processes in electron-proton and electron-heavy-ion collisions require a good understanding of the high-energy limit of QCD, in particular the parton saturation regime \cite{Gribov:1984tu,Mueller:1985wy,McLerran:1994ni,McLerran:1994ka,McLerran:1994vd,Mueller:1999wm,Gelis:2010nm,Albacete:2014fwa}. 

In the high energy limit the eikonal propagation of a colored probe is determined by the Wilson lines, the path-ordered exponentials of the hadron's color field. The color glass condensate (CGC) framework provides an effective field theory description of high-energy QCD, which can be formulated entirely in terms of the Wilson lines. The Wilson lines are treated as stochastic variables and their distribution characterizes the properties of the hadronic target when probed at a given energy.

The energy dependence of this distribution is described by the JIMWLK renormalization group equation \cite{Balitsky:1995ub,Jalilian-Marian:1997jx,*Jalilian-Marian:1997gr,*Iancu:2000hn,*Ferreiro:2001qy,*Mueller:2001uk}. It can be derived by successively integrating out quantum fluctuations at lower and lower Bjorken $x$ and including them in the effective theory by renormalizing the statistical distribution.

The JIMWLK equation was formulated as a non-linear stochastic process in \cite{Weigert:2000gi}. This functional Langevin form provides an intuitive picture of the physical processes involved as one evolves towards higher energy, and is particularly well suited for numerical implementations. 
First numerical solutions of the JIMWLK equation were presented in \cite{Rummukainen:2003ns,Kovchegov:2008mk}. Further numerical studies including running coupling effects were conducted in \cite{Lappi:2011ju,Dumitru:2011vk,Lappi:2012vw}. 

So far, all explicit solutions of the JIMWLK equation have assumed infinitely large and homogeneous nuclei. 
In this work we introduce for the first time an explicit impact parameter dependence and study the small $x$ evolution of the gluon distribution in a
finite size proton. This will allow for the calculation of the spatial substructure of the proton at high energies, including fluctuations in coordinate space, which are potentially of great importance for the interpretation of recent RHIC and LHC results on p/d-A and high-multiplicity p-p collisions \cite{Khachatryan:2010gv,Adare:2013piz,CMS:2012qk,Abelev:2012ola,Aad:2012gla}.

Generally, for finite size nuclei, a particular problem of the JIMWLK equation
becomes apparent - due to the presence of long range Coulomb fields, the size of the nucleus will grow exponentially as we evolve towards small $x$ \cite{Kovner:2001bh,GolecBiernat:2003ym,Berger:2010sh}. This has dramatic consequences because it leads to a violation of unitarity and confinement. The Froissart bound \cite{Froissart:1961ux,Martin:1962rt} requires that the total inelastic cross section for the scattering of two hadrons fulfills
\begin{equation}\label{eq:Froissart}
  \sigma < \pi d^2 \ln^2(s/s_0)\,,
\end{equation}
where $d$ is a typical hadronic size scale, $s$ is the center of mass energy squared, and $y=\ln(s/s_0)$ is the rapidity. 
This relation is fulfilled only when the energy dependence of the hadron radius $R_{\rm h}$ is $\sim d \ln(s/s_0)$ or weaker.

As discussed in Refs.\,\cite{Kovner:2001bh,GolecBiernat:2003ym}, the apparent problem with the JIMWLK equation is the absence of confinement effects in the evolution kernel. While the (perturbative) JIMWLK kernel allows for successive gluon radiation at large distance scales, confinement effects should suppress such emissions and slow down the growth of the hadron.

In this work we will account for confinement effects on a phenomenological level by introducing a modification of the JIMWLK kernel, which leads to an exponential suppression of large distance gluon emission via an effective mass term. In this case the proton grows only linearly with rapidity, respecting the Froissart bound. 
We discuss effects of the modification on the evolution of the saturation scale and the effective proton radius.

For the description of a proton at high energies, we model the initial configuration at moderately small $x$ by assuming that gluon distributions are concentrated around sampled valence quark positions. We then study how JIMWLK evolution modifies the transverse spatial structure of the gluon distribution. We quantify the proton size and the saturation scale, as well as their evolution speeds, via the dipole scattering amplitude.

This letter is organized as follows. In Section \ref{sec:langevin} we introduce the Langevin form of the JIMWLK equation and the modification of the kernel used to regulate the infrared physics. In Section \ref{sec:numerics} we discuss details of the numerical implementation. We present numerical results on the spatial structure of a single proton in Section \ref{sec:SingleProton}, discuss features of the dipole scattering amplitude in Section \ref{sec:DipoleAmplitude}, and extract the saturation scale and proton radius as a function of rapidity in Section \ref{sec:Unitarity}. We conclude in Section \ref{sec:Conclusion}.

\section{Langevin form of the JIMWLK equation}\label{sec:langevin}
The evolution equation of the probability distribution of the Wilson lines to leading logarithmic accuracy ($\alpha_s\ln(1/x)$) can be written as a functional Fokker-Planck equation \cite{Weigert:2000gi}. 
It can be re-expressed as a functional Langevin equation for the Wilson lines themselves \cite{Blaizot:2002xy}:
\begin{equation}\label{eq:langevin}
  \frac{d}{dY} V_{\mathbf{x}} = V_{\mathbf{x}} (i t^a) \Big [ \int_{\mathbf{z}} \varepsilon_{\mathbf{x},\mathbf{z}}^{a b, i}\, \xi_{\mathbf{z}, i}^b(Y)+\sigma_{\mathbf{x}}^a\Big]\,,
\end{equation}
where $\mathbf{x}$ and $\mathbf{z}$ are two-dimensional vectors in the transverse plane. The Wilson line $V$ is a unitary matrix and $t^a$ are the $SU(N_{\rm c})$ generators in the fundamental representation. $i=1,2$ is a transverse spatial index, and $a,b \in \{1,\dots,N_c^2-1\}$  are the color indices.
We used the shorthand notation $\int_{\mathbf{z}}=\int d^2z$.

The terms in square brackets in (\ref{eq:langevin}) can be interpreted as a stochastic random noise term and a deterministic drift term.
The random noise is Gaussian and local in transverse coordinate, color, and rapidity: $\langle \xi_{\mathbf{z}, i}^b(Y)\rangle = 0$ and
\begin{equation}\label{eq:noise}
  \langle \xi_{\mathbf{x}, i}^a(Y) \xi_{\mathbf{y}, j}^b(Y')\rangle = \delta^{ab}\delta^{ij}\delta_{ \mathbf{x}\mathbf{y} }^{(2)} \delta(Y-Y')\,.
\end{equation}

The coefficient of the random noise is given by
\begin{equation}
  \varepsilon_{\mathbf{x},\mathbf{z}}^{a b, i} = \left(\frac{\alpha_s}{\pi^2}\right)^{1/2} K^i_{\mathbf{x}-\mathbf{z}}\left[1-U^\dag_{\mathbf{x}}U_{\mathbf{z}}\right]^{ab}\,,
\end{equation}
where $U$ is the Wilson line in the adjoint representation, and the vector kernel is
\begin{equation}\label{eq:K}
  K^i_{\mathbf{r}} = \mathbf{r}^i/\mathbf{r}^2\,.
\end{equation}

The ``drift term'' is given by 
\begin{equation}
  \sigma_{\mathbf{x}}^a = -i \frac{\alpha_s}{2\pi^2}\int_{\mathbf{x}}S_{\mathbf{x}-\mathbf{z}}\, \tr [T^a U^\dag_{\mathbf{x}}U_{\mathbf{z}}]\,,
\end{equation}
with the scalar kernel  $S_{\mathbf{r}} = 1/\mathbf{r}^{2}$ and $T^a$ a generator in the adjoint representation.

In the numerical solution of Eq.\,(\ref{eq:langevin}) a discrete rapidity step will be employed. For one step of size $dY$, the change 
in the Wilson line is given by
\begin{equation}\label{eq:step}
  V_{\mathbf{x}}(Y+dY) = V_{\mathbf{x}}(Y) \exp \Big\{ i t^a \int_{\mathbf{z}} \varepsilon_{\mathbf{x},\mathbf{z}}^{a b, i}\, \xi_{\mathbf{z}, i}^b \sqrt{dY}+\sigma_{\mathbf{x}}^a dY\Big\}\,.
\end{equation}
The delta-function in rapidity $\delta(Y-Y')$ in (\ref{eq:noise}) becomes a Kronecker-Delta divided by the magnitude of the timestep:
$\delta(Y_m-Y_n) \rightarrow \delta_{m,n}/dY$, which is removed from the normalization of the noise. This leads to the appearance of $\sqrt{dY}$ in the first term in the exponential in Eq.\,(\ref{eq:step}).

In \cite{Lappi:2012vw} the following simpler form of the Langevin step was derived:
\begin{align}\label{eq:newstep}
  V_{\mathbf{x}}(Y+dY) = \exp \Big\{-i \frac{\sqrt{\alpha_s dY}}{\pi} \int_{\mathbf{z}} \mathbf{K}_{\mathbf{x}-\mathbf{z}}\cdot (V_{\mathbf{z}}\boldsymbol{\xi}_{\mathbf{z}}
V^\dag_{\mathbf{z}})\Big\}\notag\\ \times V_{\mathbf{x}}(Y) \exp \Big\{ i \frac{\sqrt{\alpha_s dY}}{\pi} \int_{\mathbf{z}} \mathbf{K}_{\mathbf{x}-\mathbf{z}}\cdot \boldsymbol{\xi}_{\mathbf{z}}\Big\}\,,
\end{align}
where $\boldsymbol{\xi}_{\mathbf{z}} = (\xi_{\mathbf{z},1}^a t^a,\xi_{\mathbf{z},2}^a t^a)$.
 By allowing multiplication of the Wilson line from the left and right, the rapidity step can be written with only a stochastic term. As noted in \cite{Lappi:2012vw} this makes the numerical evaluation significantly more efficient, since no adjoint Wilson lines need to be computed.
In the limit $dY\rightarrow 0$ the update step (\ref{eq:newstep}) is exactly equivalent to the original one (\ref{eq:step}). The difference appears at order $\mathcal{O}(dY^{3/2})$, which has been neglected in the derivation of both expressions \cite{Blaizot:2002xy,Lappi:2012vw}.

As discussed above, the rapidity evolution will lead to exponential growth of the total inelastic cross section with rapidity. To tame this behavior we need to model the effect of confinement and constrain the kernel (\ref{eq:K}) at large distance scales.
A simple method is to include an exponential screening as already suggested in \cite{Kovner:2001bh}. We implement this in practice by introducing an effective mass scale $m$ on the order of $\Lambda_{\rm QCD}$ which modifies the kernel according to
\begin{equation}\label{eq:modkernel}
  \mathbf{K}^{\text{(mod)}}_{\mathbf{r}} = m|\mathbf{r}|~K_{1}(m |\mathbf{r}|)~\mathbf{K}_{\mathbf{r}}.
\end{equation}
Here $K_{1}(x)$ is the modified Bessel function of the second kind. The limiting behavior is such that  at small arguments $x K_{1}(x)=1+\mathcal{O}(x^2)$  and no modifications of the kernel occur, whereas  for large arguments  $K_{1}(x)=\sqrt{\pi/(2x)} e^{-x}$ decays exponentially.
This modification breaks gauge invariance, however, it allows for a systematic analysis of the effect of the mass term. We will discuss the dependence of the saturation scale and the proton radius on its value below.

\begin{figure*}[t!]
\includegraphics[width=\textwidth]{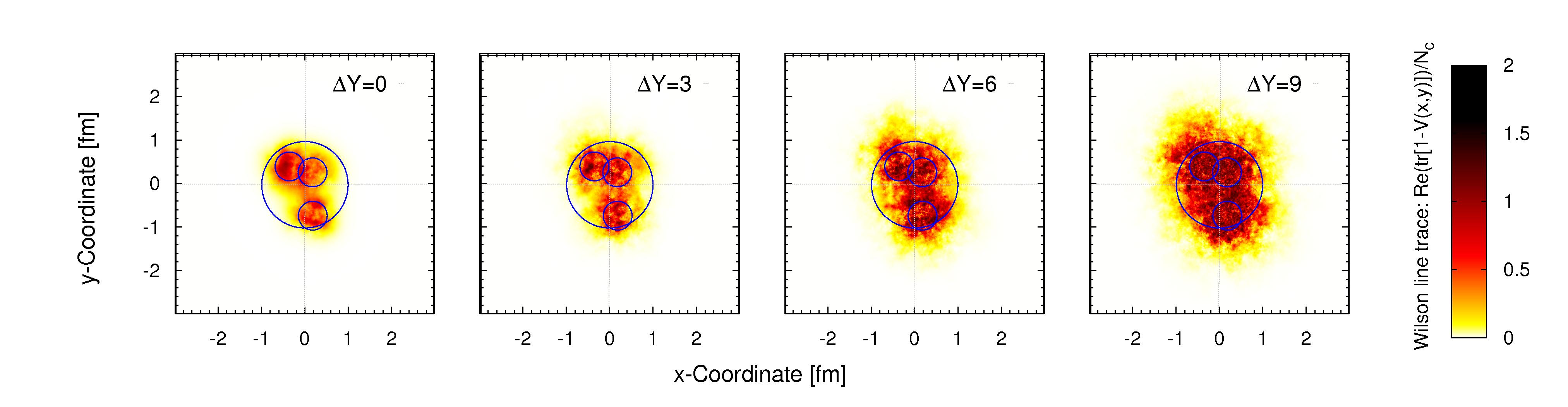}

\caption{
Transverse profile of a single proton configuration at four different intervals $dY$ of the evolution. The different panels show a contour plot of the real part of the trace of the Wilson line $\Re(\tr[\id-V(x,y)])/\nc$ as a function of the transverse coordinates $x$ and $y$. The small (large) blue circles show the position and size of the three constituent quarks (the proton).\label{fig:tracemap}
}
\end{figure*}

\section{Numerical implementation}\label{sec:numerics}
We follow previous works \cite{Rummukainen:2003ns,Lappi:2011ju,Dumitru:2011vk,Lappi:2012vw} in the numerical implementation of the JIMWLK equation and first discretize the transverse space on a spatial lattice with $\Nt \times \Nt$ points, where adjacent lattice points are separated by the lattice spacing $\at$. The Wilson lines $V_{\mathbf{x}}$ as well as the stochastic fields $\boldsymbol{\xi}_{\mathbf{x}}$ are defined on the points $\mathbf{x}=(x_{1},x_{2})$ with $x_{1/2}=0,\cdots,\Nt-1$  of the transverse lattice.  We employ the formulation of the JIMWLK equation in $(\ref{eq:newstep})$, which only involves the vector kernel $\mathbf{K}$.

We employ periodic boundary conditions in the transverse plane for all dynamical fields, which as discussed previously \cite{Rummukainen:2003ns} greatly reduces the computational expense, because Fourier acceleration can be employed. We emphasize that even though we will be interested in the evolution of a finite size proton where translation invariance is explicitly broken, the use of periodic boundary conditions does not pose any additional problems. We find that, as long as the kernel decays sufficiently fast at large distance scales and the physical extent of the proton is small compared to the lattice size, (unphysical) contributions from across the lattice boundary are suppressed by several orders of magnitude.

We solve the lattice version of Eq.\,(\ref{eq:newstep}) numerically by performing a series of updates in $dY$ according to the following procedure: We first generate the stochastic fields $\boldsymbol{\xi}_{\mathbf{x}}$ at each lattice point and subsequently perform the color rotations  $V_{\mathbf{z}}\boldsymbol{\xi}_{\mathbf{z}}V^\dag_{\mathbf{z}}$ to obtain the argument of the left hand side exponential.  We then perform the two convolutions with the kernel, which for a lattice with  periodic boundary conditions can be performed in Fourier space at cost of order $\Nt^2 \log(\Nt^2)$, which is significantly more efficient than the direct implementation in coordinate space, which scales as $\Nt^4$. Finally, we perform the matrix exponential by use of analytic matrix diagonalization formulae \cite{Kopp:2006wp} and compute the Wilson lines at the next rapidity step. This procedure is then repeated to obtain the evolution over a finite rapidity interval.

Within this framework observables can be computed in a straightforward way as functionals of the Wilson lines at any given $Y$ \cite{Rummukainen:2003ns,Lappi:2011ju,Dumitru:2011vk,Lappi:2012vw}. When converting the results to physical units, the scale of the lattice computation is set by the proton radius $R_{\rm p}\simeq 1~{\rm fm}$.\footnote{The precise value of $R_{\rm p}$ can be fixed by fitting experimental data on DIS cross sections within our model. We expect $R_{\rm p}$ to be close to the gluonic radius (see e.g. \cite{Caldwell:2010zza}).}  If not stated otherwise, the results presented in this paper are obtained for $\Nt=2048$ lattices with physical size $\Nt\at=8.53~{\rm fm}$, lattice spacing $\at=4.167 \times 10^{-3}~{\rm fm}$, and rapidity step size $dY=3.33 \times 10^{-3}$. We will consider a fixed coupling constant $\alpha_s=0.3$ for simplicity and comment on expected modifications due to running coupling effects.

\begin{figure*}[t!]
\includegraphics[width=\textwidth]{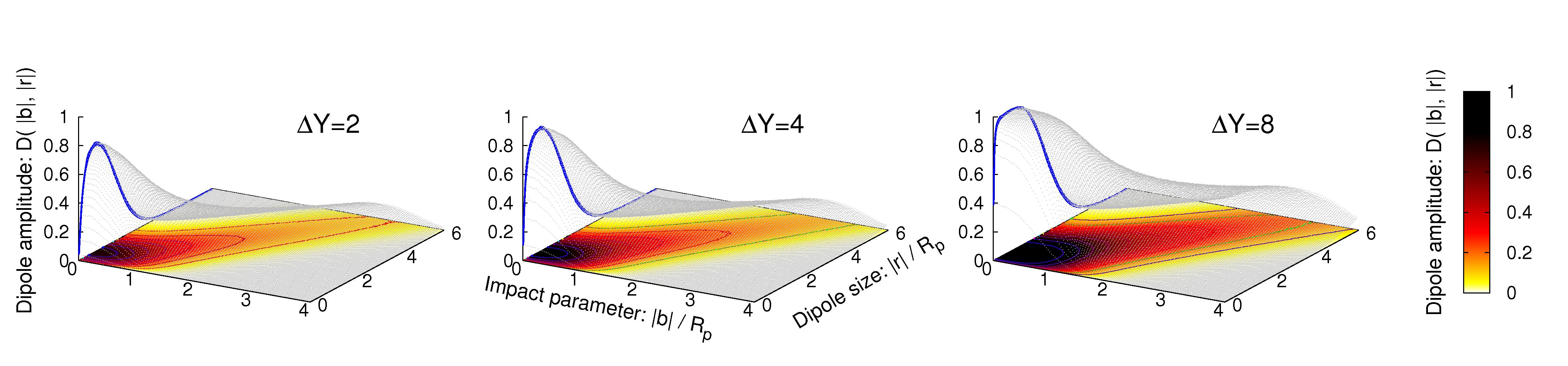}
\caption{
Dipole scattering amplitude as a function of impact parameter $|\mathbf{b}|$ and dipole size $|\mathbf{r}|$ measured at three different intervals $\Delta Y$ of the evolution. While the dominant support is at small impact parameter $|\mathbf{b}|\lesssim R_{\rm p}$ for hadronic size dipoles $|\mathbf{r}|\simeq R_{\rm p}$, one also observes a band around $|\mathbf{r}|\simeq 2 |\mathbf{b}|$ for larger impact parameters. \label{fig:2dDipole}
}
\end{figure*}

\section{Evolution of a single proton}\label{sec:SingleProton}
When studying the energy evolution of a single proton, we start at some moderately small value of $x=x_{0}$, where the evolution becomes dominated by the gluon degrees of freedom. We thus need a parametrization of the initial Wilson line configurations of a proton at $x_0$, which in principle could be constrained by DIS data. Within this exploratory study, we refrain from performing actual fits to experimental data and instead consider different parameters within a simple model of the proton.

Our approach is motivated by the phenomenologically successful constituent quark model
\cite{Eremin:2003qn,Adler:2013aqf} and amounts to sampling a distribution of moderately small $x$ gluons around the large $x$ constituent quarks. In practice we first sample the positions $\vec{x}_{\rm CQ}=(\mathbf{x}_{\rm CQ},z_{\rm CQ})$ of the three large $x$ constituent quarks according to a three dimensional Gaussian distribution inside the proton, such that
\begin{eqnarray}
\langle\vec{x}_{\rm CQ}^{~2}\rangle=R_{\rm p}^{2}\,.
\end{eqnarray}
We then initialize the Wilson lines according to a color neutral distribution of randomly distributed color charges $\rho_{a}(\mathbf{x})$ inside the constituent quarks,  which we think of as corresponding to the gluons radiated  off the constituent quarks between $x\sim1$ and the initial value of $x=x_{0}$. 

We divide this large $x$ region into $N^{0}_{Y}=100$ intervals, such that the initial Wilson lines are given by \cite{Lappi:2007ku}
\begin{eqnarray}\label{eq:Wilson}
V_{0}(\mathbf{x})= \prod_{i=1}^{N^{0}_{Y}} \exp \left( -i g \frac{\rho^{Y_{i}}_{a}(\mathbf{x}) t^{a}}{\boldsymbol{\nabla}_\perp^2+m^2} \right)
\end{eqnarray}
where $\boldsymbol{\nabla}_\perp^2 = \partial_i\partial^i$ and $m\sim\Lambda_{QCD}$ is the same effective mass scale as in Eq.\,(\ref{eq:modkernel}), which regulates the infrared behavior of the Coulomb tails. We consider a Gaussian distribution of the color charges $\rho_{a}^{Y_{i}}(\mathbf{x})$, which  -- following standard McLerran-Venugopalan type models \cite{McLerran:1994ni} -- we take as uncorrelated between points ($\mathbf{x}$ and $\mathbf{y}$) in the transverse plane, different colors, and different rapidity intervals ($Y_{i}$ and $Y_{j}$), i.e.,
\begin{eqnarray}\label{eq:rhorho}
g^2 \langle\rho_{a}^{Y_{i}}(\mathbf{x})\rho_{b}^{Y_{j}}(\mathbf{y})\rangle &=& \frac{\left(g^2\mu_{0}R_{\rm CQ}\right)^2}{N^{0}_{Y}}~S\left(\frac{\mathbf{x+y}}{2}\right) \\
&&\times~\delta_{ab}~\delta_{Y_{i} Y_{j}}~\delta^{(2)}(\mathbf{x-y})\;.\nonumber 
\end{eqnarray}
The spatial distribution of the color charge density  $S\left(\frac{\mathbf{x+y}}{2}\right)$ is centered around the constituent quarks according to
\begin{eqnarray}
S(\mathbf{x}) = \frac{3}{2 \pi R^{2}_{\rm CQ}} \sum_{n=1}^{N_{\rm CQ}} \exp \left(  -\frac{3}{2 R_{\rm CQ}^{2}} \Big(\mathbf{x}-\mathbf{x}^{(n)}_{\rm CQ}\Big)^2 \right)
\end{eqnarray}
and normalized to the number of constituent quarks $N_{\rm CQ}=\int d^2\mathbf{x}~S(\mathbf{x})=3$.
The radius of the "gluon cloud" around the constituent quark is set to $R_{\rm CQ}=R_{\rm p}/3$ in the following. We will vary the confinement scale $m$ in (\ref{eq:modkernel}) and (\ref{eq:Wilson}) as well as the dimensionless parameter $g^2\mu_{0}R_{\rm CQ}$ in (\ref{eq:rhorho}), which controls the initial degree of non-linearity of the subsequent small $x$ evolution.\footnote{We find that for typical values of $m$ considered here, values of  $g^{2}\mu_{0}R_{\rm CQ}\sim 1$ or less initially correspond to a linear evolution (BFKL regime), whereas non-linear evolution effects and gluon saturation are immediately important for $g^{2}\mu_{0}R_{\rm CQ}\sim 10$ or greater.}  If not stated otherwise results are presented for $mR_{\rm p}=3$ and $g^2\mu R_{\rm p}=30$.

We first study the small $x$ evolution of the spatial sub-structure of a single proton. In Fig.\,\ref{fig:tracemap}, we present the result for $\Re(\tr[\id-V(\mathbf{x})])/\nc$ in a single configuration at different rapidity intervals. This quantity is the simplest way to characterize the distribution of the gluon field in transverse space.
The proton's position and average initial radius is given by the larger circle while initial positions of ``constituent quarks'' are
marked by the smaller circles of radius $R_{\rm p}/3$ in all plots. 

The structure of the initial gluon distribution ($\Delta Y=0$) is dominated by the three ``constituent quark'' positions but also shows additional gluon field fluctuations. We observe that after evolution to $\Delta Y=3$, 6, and even 9, there is still a noticeable imprint of the larger $x$ structure on the gluon field distribution. 

This result demonstrates for the first time that even at very high energies and rapidities, the shape of the gluon distribution in a proton can fluctuate significantly. This can have important effects on multi-particle production and correlations in high-multiplicity p+p and p+heavy-ion collisions. Both initial state correlations as discussed in \cite{Dumitru:2010iy,Dusling:2012iga,Dusling:2012cg,Dusling:2012wy,Dusling:2013oia} and final state collective effects (see \cite{Schenke:2014zha}) can be strongly affected by the fluctuating structure of the gluon distribution. 

\begin{figure*}[ht]
\begin{minipage}[t]{0.45\linewidth}
\centering

\includegraphics[width=\textwidth]{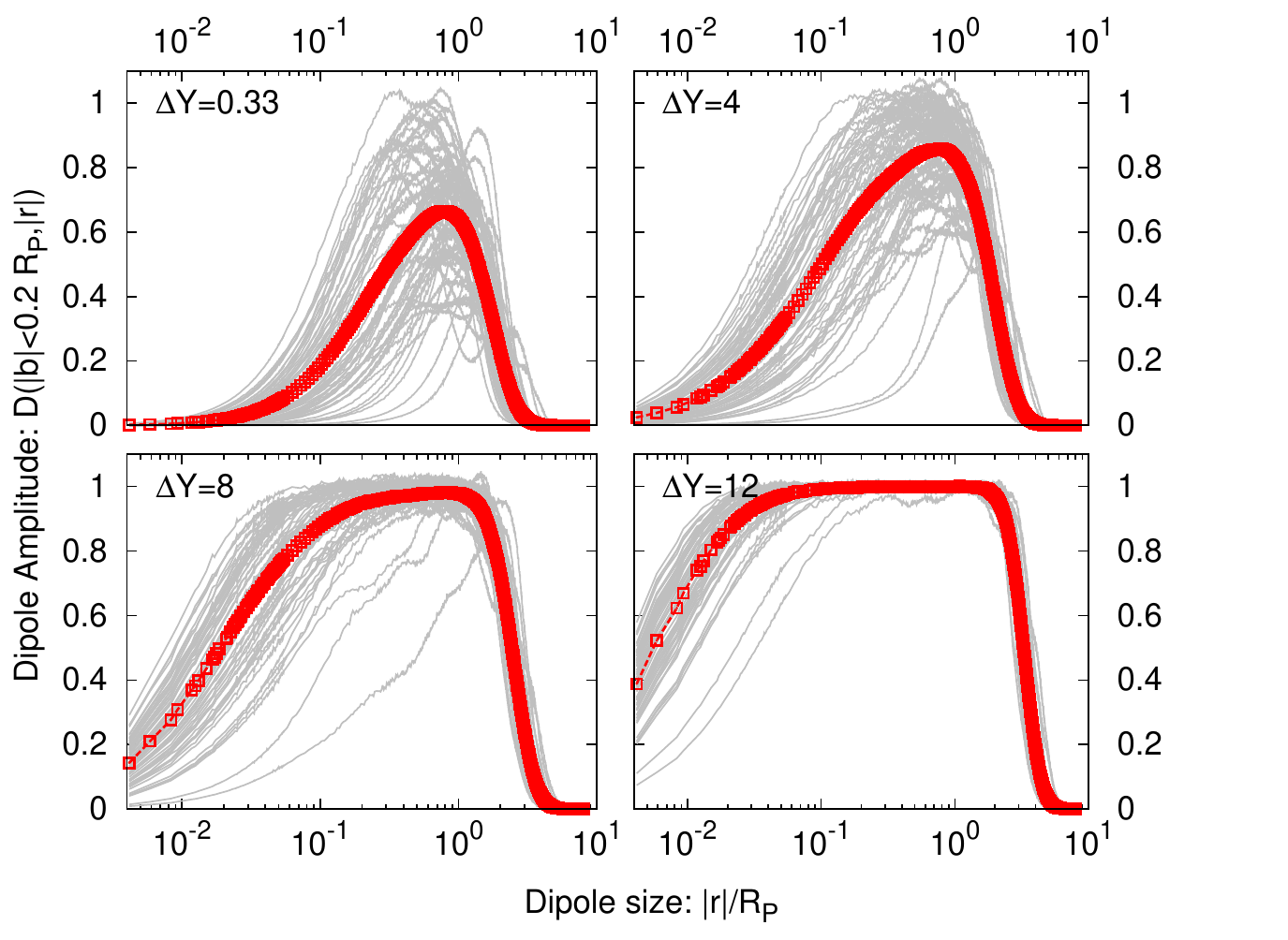}
\caption{
Dipole scattering amplitude as a function of dipole size $|\mathbf{r}|$ at fixed impact parameter $|\mathbf{b}|<0.2\, R_{\rm p}$ measured at four different intervals $\Delta Y$ of the evolution. The gray lines show the result for $N_{\text{conf}}=64$ different initial configurations and illustrate the magnitude of event-by-event fluctuations. The red points correspond to the average over initial configurations.\label{fig:Dr}
}

\end{minipage}
\hspace{0.5cm}
\begin{minipage}[t]{0.45\linewidth}
\includegraphics[width=\textwidth]{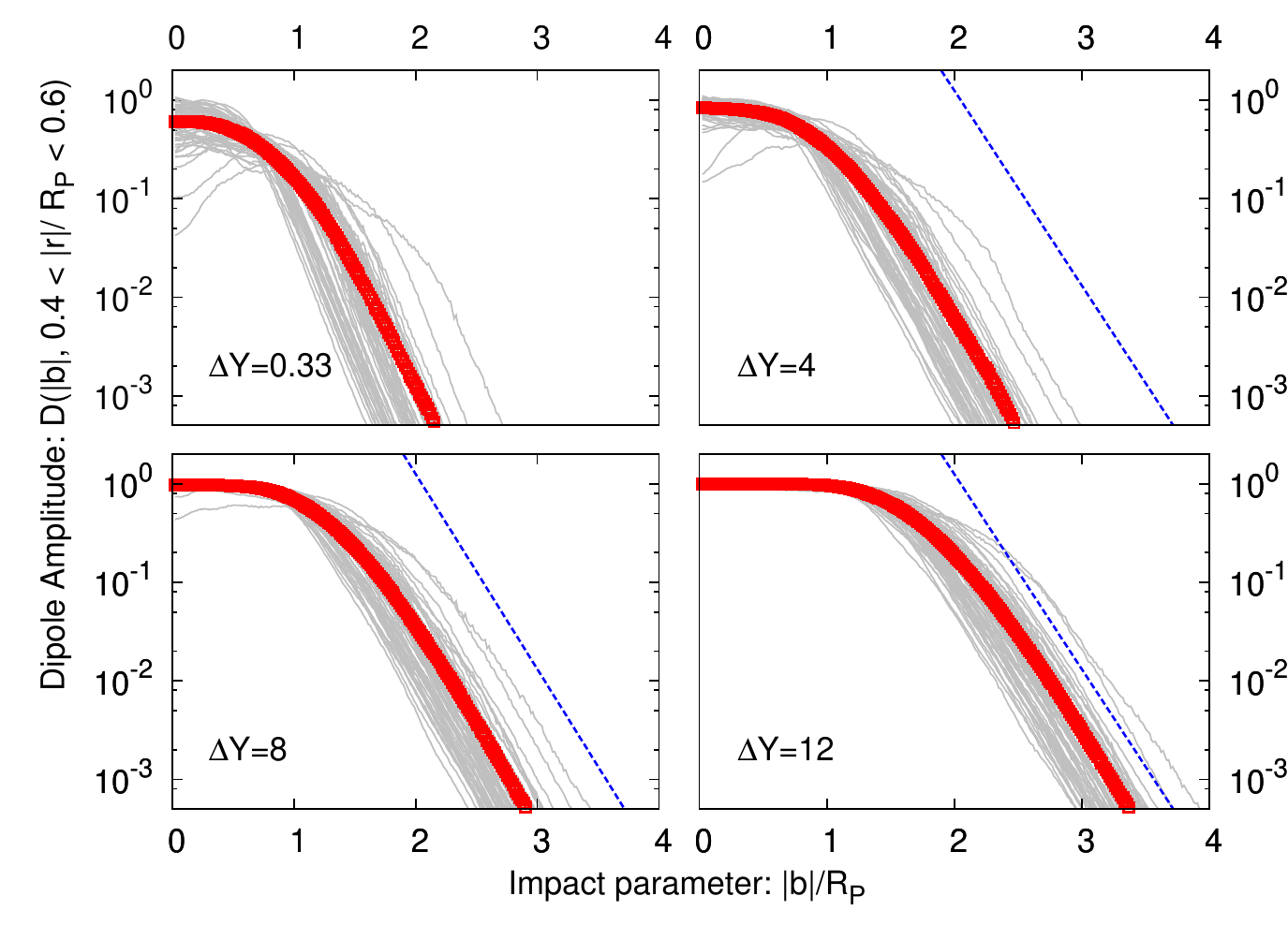}
\caption{
Dipole scattering amplitude as a function of impact parameter $|\mathbf{b}|$ for a hadronic size dipole with $0.4<|\mathbf{r}|/R_{\rm p}<0.6$ measured at four different intervals $\Delta Y$ of the evolution. The gray lines show the result for $N_{\text{conf}}=64$ different initial configurations. The red points correspond to the average over initial configurations. The blue dashed line illustrates the exponential decay of the Dipole amplitude. \label{fig:Db}
}

\end{minipage}
\end{figure*}

\section{Dipole scattering amplitude}\label{sec:DipoleAmplitude}
A natural way to characterize the gluon distribution of the proton is via the scattering amplitude of a color singlet dipole with charges at points $\mathbf{x}$ and $\mathbf{y}$
\begin{equation}
  D(\mathbf{x},\mathbf{y}) = \frac{1}{\nc} \tr \langle 1 - V^\dag(\mathbf{x}) V(\mathbf{y})\rangle\,.
\end{equation}
In the following we will express the dipole amplitude $D$ as a function of the impact parameter $\mathbf{b}=(\mathbf{x}+\mathbf{y})/2$, measured relative to the center of mass of the constituent quarks, and the dipole size (and orientation) $\mathbf{r}=\mathbf{x}-\mathbf{y}$.

The dipole scattering amplitude averaged over $N_{{\rm conf}}=64$ configurations and the orientations of $\mathbf{b}$ and $\mathbf{r}$, $D(|\mathbf{b}|, |\mathbf{r}|)$, is shown in Fig.\,\ref{fig:2dDipole} with $|\mathbf{b}|$ and $|\mathbf{r}|$ in units of the proton radius $R_{\rm p}$.
Three different rapidity values are shown in three separate plots.

The strongest support of the dipole amplitude resides in the region of small impact parameters $|\mathbf{b}|$, close to the center of the proton. The behavior at small values of $|\mathbf{r}|$ is similar to the case of infinite nuclei. Starting from $|\mathbf{r}|=0$, where the dipole amplitude vanishes by definition, one observes a rise towards larger values of $|\mathbf{r}|$, where the dipole amplitude reaches a maximum for $|\mathbf{r}|\sim R_{\rm p}$. While initially at $\Delta Y=0$, this maximum value is below unity, it quickly approaches the saturation bound $(D=1)$ as the rapidity evolution proceeds.

The drop of the dipole amplitude for values of the dipole size $|\mathbf{r}|\gtrsim R_{\rm p}$ (and small $|\mathbf{b}|$)
occurs when the points $\mathbf{x}$ and $\mathbf{y}$ both fall outside the effective proton radius, i.e., outside the region of coordinate space where the Wilson lines are significantly different from unity. Further, we find that for dipoles of size $|\mathbf{r}|\gtrsim R_{\rm p}$, the dipole amplitude is maximal around $|\mathbf{r}|\simeq 2 |\mathbf{b}|$. This has been observed previously in \cite{GolecBiernat:2003ym} and can be easily understood, as in this configuration there are certain angular orientations for which one end of the dipole lies in the center of the proton, where on average $V$ differs maximally from unity. 

We note that confinement effects that are not included here will modify the behavior of the dipole amplitude at large $|\mathbf{r}|$. This is because in the full theory new quark -- anti-quark pairs will be formed as one increases the dipole size $|\mathbf{r}|$ beyond the size scale $1/\Lambda_{\rm QCD}$. At small impact parameter $|\mathbf{b}|$ this effect will lead to large scattering cross sections even at large values of $|\mathbf{r}|$ and cause the full dipole amplitude to remain close to unity. We leave the inclusion of this effect to future work, where cross sections for electron-proton scattering will be computed.

In Fig.\,\ref{fig:Dr} we show the dipole scattering amplitude $D$ as a function of dipole size $|\mathbf{r}|$ at fixed impact parameter $|\mathbf{b}|<0.2 R_{\rm p}$, measured at four different intervals $\Delta Y$ of the evolution. We show the variation of $D$ configuration-by-configuration by plotting as narrow lines results for 64 individual initial configurations. The average is shown by the red points. We note that the variance of the distribution is largest at the smallest $\Delta Y$. The fluctuations are dominated by the positions of the ``constituent quarks'' that may or may not be located at $|\mathbf{b}|<0.2\, R_{\rm p}$ in a given configuration. Differences between different configurations are still large at $\Delta Y=4$ and even  $\Delta Y=8$, in line with our findings in Fig.\,\ref{fig:tracemap}. Only as we evolve towards very large $\Delta Y$ the variance is reduced, indicating that gluon distributions in the proton at very small $x$ are more universal. Fluctuations of the dipole amplitude then translate into fluctuations of the typical momentum scale $Q$, which can have important effects for the calculation of observables in hadronic collisions \cite{Schenke:2013dpa}.

We note again that the drop of the dipole scattering amplitude at large $|\mathbf{r}|\gtrsim R_{\rm p}$ happens because we allow a single dipole to stretch larger than typical hadronic sizes. In a complete, non-perturbative theory this can not happen -- confinement effects would cause pair production, strongly modifying the dipole cross section at large $|\mathbf{r}|$.

Fig.\,\ref{fig:Db} shows the dipole scattering amplitude as a function of impact parameter $|\mathbf{b}|$ for a dipole of hadronic size $0.4<|\mathbf{r}|/R_{\rm p}<0.6$ measured at four different intervals $\Delta Y$ of the evolution. Again, one can see a rather wide spread around the average, generated mainly by the ``constituent quark'' fluctuations, especially at small $\Delta Y$. At larger $\Delta Y$ almost all configurations have reached saturation at small $|\mathbf{b}|$. At large $|\mathbf{b}|$ the dipole amplitude falls off exponentially, because of the exponential regulator we introduced in Eqs.\,(\ref{eq:modkernel}) and (\ref{eq:Wilson}).
The growth of the proton with increasing $\Delta Y$ is nicely visible. We will analyze this growth in more detail in the following section.

\begin{figure*}[ht]
\begin{minipage}[t]{0.45\linewidth}
\centering

\includegraphics[width=\textwidth]{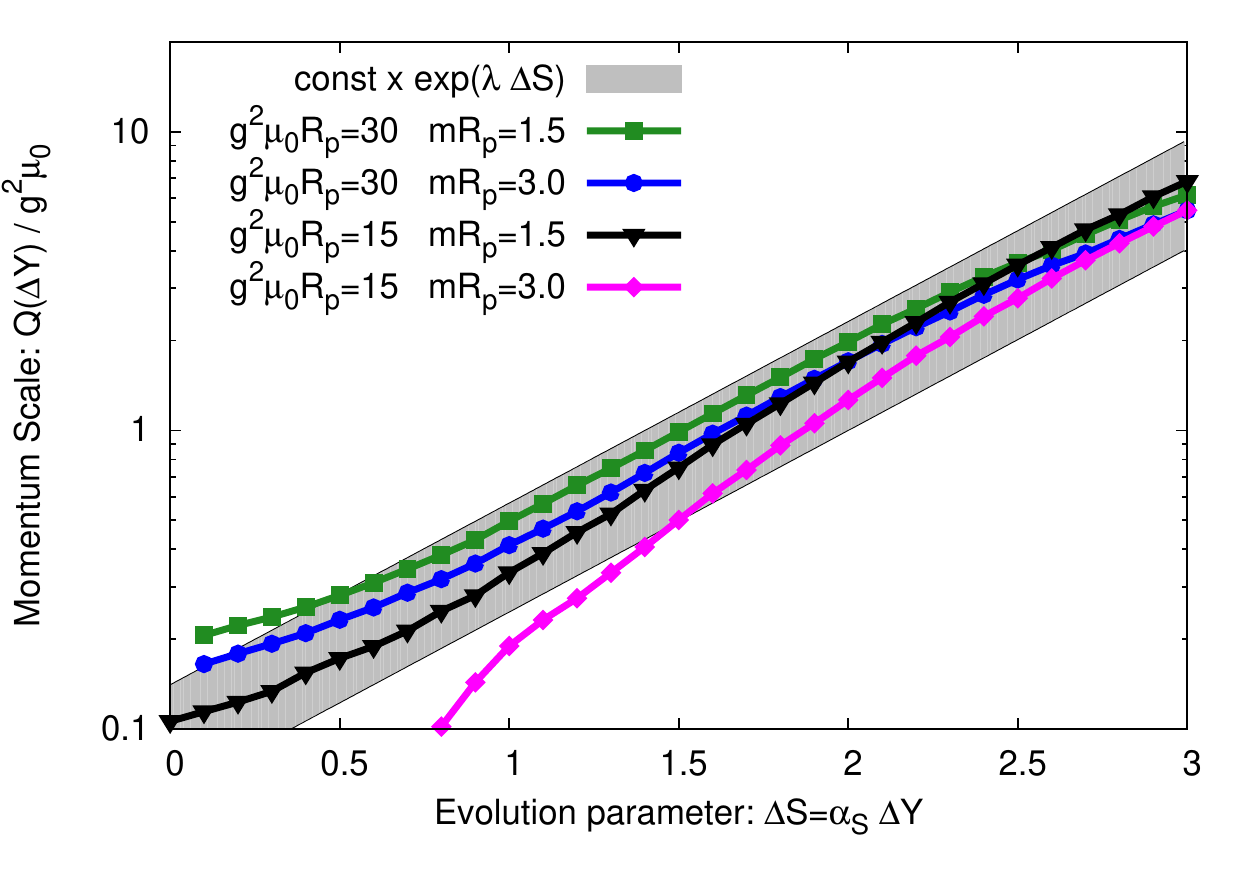}
\caption{
Characteristic momentum scale $Q$ as a function of the ultra-violet evolution variable $\Delta S=\alpha_{s} \Delta Y$ for different sets of parameters $g^{2}\mu_{0}R_{\rm p}$ and $mR_{\rm p}$. The gray band illustrates the \textit{\textbf{exponential growth}} of $Q(\Delta Y)/Q_{0}\propto \exp(\lambda \alpha_{s} \Delta Y)$ with $\lambda\simeq 1.4$. \label{fig:Qs}}

\end{minipage}
\hspace{0.5cm}
\begin{minipage}[t]{0.45\linewidth}
\includegraphics[width=\textwidth]{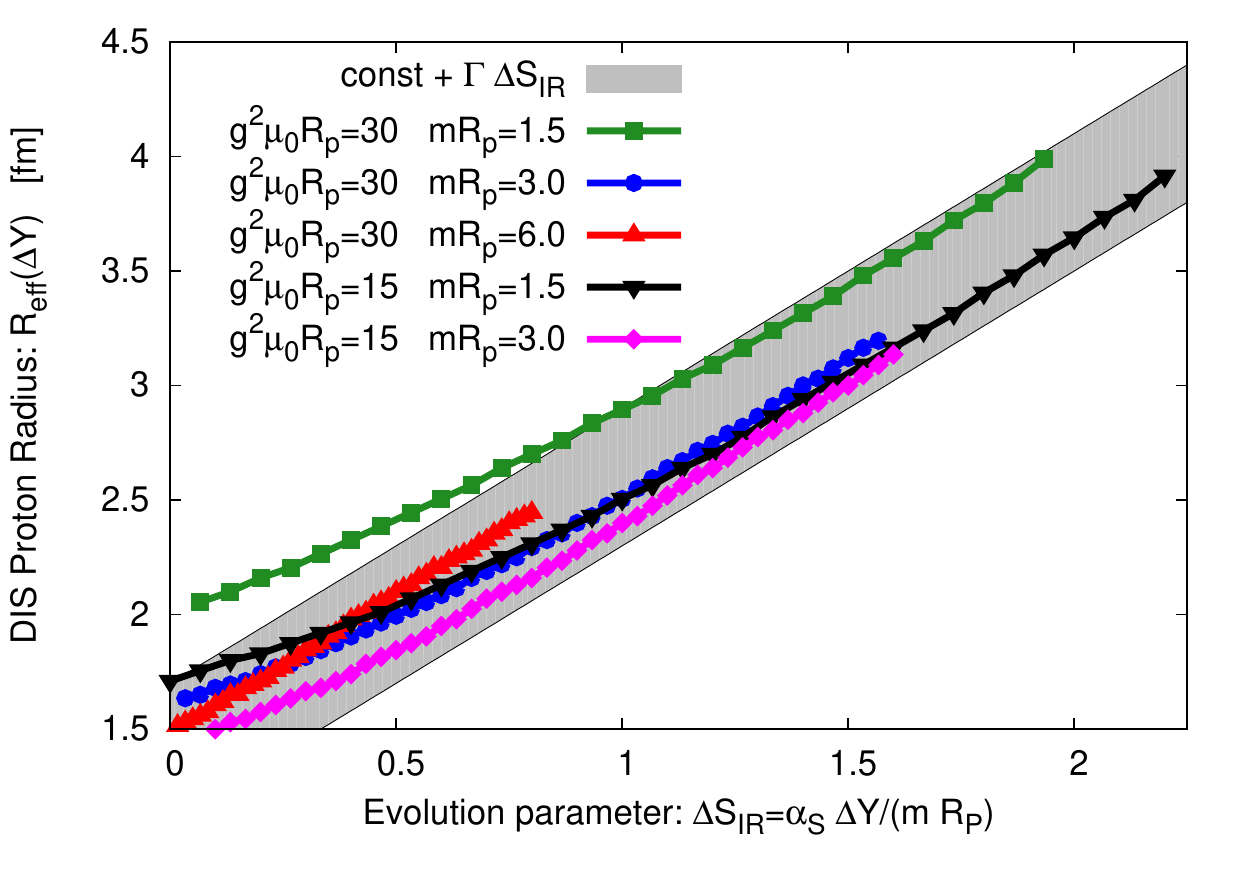}

\caption{
Effective proton radius $R_{\text{eff}}$ as a function of the infrared evolution variable $\Delta S_{IR}=\alpha_{s} \Delta Y/(m R_{\rm p})$ for different sets of parameters $g^{2}\mu_{0}R_{\rm p}$ and $mR_{\rm p}$. The gray band illustrates the \textit{\textbf{linear growth}} of $R_{\text{eff}}(\Delta Y)/R_{\rm p}\propto  \Gamma \alpha_{s} \Delta Y / (m R_{\rm p})$ with $\Gamma\simeq 1.2$. \label{fig:r}
}
\end{minipage}
\end{figure*}

\section{Saturation \& Unitarity}\label{sec:Unitarity}
We have demonstrated in the previous section that saturation is reached for impact parameters $|\mathbf{b}|\lesssim R_{\rm p}$ and large enough $\Delta Y$.
Here we analyze the growth of the typical momentum scale $Q$, defined via
\begin{equation}
  D(|\mathbf{b}|<0.2 R_{\rm p}, |\mathbf{r}|=1/Q) = e^{-1}\,,
\end{equation}
for various sets of parameters. 
We chose a smaller reference value $D(|\mathbf{r}|=1/Q)=e^{-1}$ than the usual $D(|\mathbf{r}|=\sqrt{2}/Q_s)=e^{-1/2}$ \cite{Kowalski:2003hm,Lappi:2011ju}, to be able to define a characteristic scale $Q$ even in situations where saturation is not yet reached. Nevertheless, $Q$ can be generally interpreted as the saturation scale $Q_s$ when considering that confinement effects will lead to $D\rightarrow 1$ as $|\mathbf{r}|\rightarrow \infty$ at any value of $\Delta Y$, as discussed above.

In Fig.\,\ref{fig:Qs} we demonstrate the exponential growth of $Q \propto \exp(\lambda \alpha_s \Delta Y)$ and extract the exponent $\lambda \simeq 1.4$, which is largely independent of the value of the confinement scale $m$ since the high-energy evolution of the saturation scale is governed by short distance physics. 
This value of $\lambda$ is consistent with the result reported in \cite{Rummukainen:2003ns} for fixed coupling JIMWLK evolution in an infinite size nucleus using similar values of $a_\bot$.

The inclusion of running coupling will slow down the growth of $Q$ according to \cite{Mueller:2002zm,Albacete:2004gw,Triantafyllopoulos:2008yn}
\begin{eqnarray}
\frac{d \log (Q^{2})}{dY} \propto \alpha_s(Q)\,.
\end{eqnarray}
This generally reduces the effective value of $\lambda$ \cite{Rummukainen:2003ns,Dumitru:2011vk}, making it more compatible with deeply inelastic scattering data \cite{GolecBiernat:1998js,Kuokkanen:2011je}.

Similarly, we can quantify the growth of the proton in impact parameter space by introducing an effective proton radius $R_{\text{eff}}$ for fixed 
$|\mathbf{r}|/R_{\rm p} \approx 0.5$ according to
\begin{equation}
  D(|\mathbf{b}|=R_{\text{eff}}, 0.4<|\mathbf{r}|/R_{\rm p}<0.6) = 0.01\,.
\end{equation}
The rapidity evolution of the effective proton radius $R_{\text{eff}}$ is shown in Fig.\,\ref{fig:r} for various sets of parameters. Since the evolution is governed by the modified kernel in Eq.\,(\ref{eq:modkernel}), which features an exponential decrease at large distance scales, we expect a linear growth of the effective proton radius in rapidity. Moreover, we expect the slope to be inversely proportional to the mass parameter \cite{Kovner:2001bh}:
\begin{equation}
\label{eq:ReffGrowth}
  R_{\text{eff}}(\Delta Y) / R_{\rm p}\propto  \Gamma \alpha_{s} \Delta Y / (m R_{\rm p})\,.
\end{equation}
When plotted as a function of the infrared evolution parameter $\alpha_{s} \Delta Y / (m R_{\rm p})$, we find that the data for different values of the parameters indeed show the same linear rise. We also determine the dimensionless slope parameter $\Gamma \simeq 1.2$, which is independent of the mass scale $m R_{\rm p}$. Moreover, since the large distance behavior is clearly controlled by the mass scale, a fixed scale of order $m$ should be the relevant scale for the coupling constant in (\ref{eq:ReffGrowth}). We therefore expect only minor modifications of the qualitative behavior in (\ref{eq:ReffGrowth}) due to running coupling effects (as long as the coupling constant is regularized in the infrared). 

As discussed above, the linear growth of the proton radius with $\Delta Y$ observed in Fig.\,\ref{fig:r} saturates the Froissart bound. The violation of unitarity due to the emission of long range Coulomb fields, which is present in the case without an infrared regulator, is therefore avoided.

\section{Conclusion}\label{sec:Conclusion}
We have presented an exploratory study of the spatial structure of the proton at high energies.
We start from a model of the large $x$ structure of the proton defining gluon distributions around constituent quark positions and determine the fixed coupling rapidity evolution from solutions to the Langevin form of the JIMWLK functional renormalization group equations. We introduce an infrared regulator that prevents the long range emission of gluons and the violation of unitarity.

An important result of our study is that the large $x$ structure of the proton affects its shape at higher energies (small $x$).
Even after evolution over $\sim 9$ units of rapidity the initial shape still affects the gluon distribution.

We presented results for the dipole scattering amplitude as a function of impact parameter and dipole size. Event-by-event fluctuations of this quantity are large at large $x$ and become smaller with decreasing $x$. We quantified effects of saturation and the growth of the proton using the dipole scattering amplitude: The saturation scale grows exponentially $Q_s \sim \exp(\lambda \alpha_s \Delta Y)$, with $\lambda \simeq 1.4$, the radius increases linearly with rapidity as $R_{\text{eff}}(\Delta Y)/R_{\rm p}\propto  \Gamma \alpha_{s} \Delta Y / (m R_{\rm p})$, with $\Gamma \simeq 1.2$.

Our findings have important implications for the physics of deeply inelastic scattering and hadronic collisions. If collective effects are relevant in high multiplicity p+A collisions, the detailed shape of the proton plays an important role. In fact, the study of azimuthal anisotropies of particle spectra produced in p+A collisions then carry information on the fluctuating shape of the proton. Here, we provide the first theoretical description of the fluctuating shape of the proton at high energies. These initial configurations can then be implemented in classical Yang-Mills simulations of the initial gluon fields in p+A collisions followed by fluid dynamic evolution as in \cite{Schenke:2014zha}.

More detailed analyses of deeply inelastic scattering observables within this model and fits to experimental data including running coupling effects are important future steps. The fluctuations included in our calculation can have important effects for the description of not only p+A and p+p collisions, but also for electron-proton collisions at current experiments and a future electron-ion collider. Even and odd azimuthal anisotropies could emerge from such a proton structure in electron-proton collisions (see e.g.\,\cite{Dumitru:2014dra}). Whether rare fluctuations as discussed in \cite{Mueller:2014fba} can be captured within our model needs to be investigated.

The presented calculations are a first step towards an event generator for electron-proton and electron-ion collisions at high collision energies.

\section*{Acknowledgments}
We thank Raju Venugopalan for very helpful discussions and comments on the manuscript. This research used resources of the National Energy Research Scientific Computing Center, which is supported by the Office of Science of the U.S. Department of Energy under Contract No. DE-AC02-05CH11231. BPS and SS are supported under DOE Contract No. DE-AC02-98CH10886. SS gratefully acknowledges a Goldhaber Distinguished Fellowship from Brookhaven Science Associates. BPS is supported by a DOE Office of Science Early Career Award.

\bibliography{spires}

\end{document}